\documentclass{conm-p-l}

\usepackage{stmaryrd,amssymb,mathrsfs}

\usepackage{tikz}
\usetikzlibrary{arrows,decorations.pathreplacing,decorations.markings,shapes.geometric,through,fit,shapes.symbols}

\tikzset{
    dot/.style={circle,draw,fill,inner sep=1pt},
    Odot/.style={circle,draw,inner sep=1pt},
    Sdot/.style={star, star point height=2pt, inner sep=.5pt, draw, fill}
}

\newcommand\define[1]{\emph{#1}}

\newcommand{\T}{{\mathrm T}}
\newcommand\KK{{\mathbb K}}
\newcommand\QQ{{\mathbb Q}}
\newcommand\RR{{\mathbb R}}
\newcommand\Ccc{{\mathscr C}}
\DeclareMathOperator{\Aut}{Aut}
\DeclareMathOperator{\ev}{ev}
\DeclareMathOperator{\homology}{H}
\renewcommand{\H}{\homology}
\newcommand\isom{\overset{\sim}{\to}}
\renewcommand{\d}{{\mathrm d}}

\newtheorem{theorem}{Theorem}[section]

\theoremstyle{definition}

\newtheorem{xca}[theorem]{Exercise}

\theoremstyle{remark}
\newtheorem{remark}[theorem]{Remark}
\newtheorem*{pref}{Prefatory remark}

\numberwithin{equation}{section}

\begin{document}

\title[Feynman diagrams from BV formalism]{How to derive Feynman diagrams for finite-dimensional integrals directly from the BV formalism}


\author{Owen Gwilliam}
\address{Department of Mathematics, Northwestern University, Evanston, IL}
\curraddr{Max Planck Institute for Mathematics, Bonn, Germany}
\email{gwilliam@mpim-bonn.mpg.de}

\author{Theo Johnson-Freyd}
\address{Department of Mathematics, University of California, Berkeley, CA}
\curraddr{Perimeter Institute for Mathematical Physics, Waterloo, Canada}
\email{tjohnsonfreyd@perimeterinsitute.ca}

\thanks{We had many rich and helpful discussions with Dan Berwick-Evans, Kevin Costello, Nicolai Reshetikhin, Thel Seraphim, and Yuan Shen about this material and many related questions.  We learned the $p$-adic remark at the end of Section~2 from Boris Hanin.  Philsang Yoo provided detailed comments on a draft of this paper.  This work was supported by NSF grants DMS-0636646 and DMS-0901431.  The original work was completed while the second author was a visitor at Northwestern University, which he thanks for its hospitality. We wish to thank an attentive, thoughtful referee for advice leading to this improved version.}

\subjclass[2000]{Primary 81S40. Secondary 18G40}


\begin{abstract}
The Batalin-Vilkovisky formalism in quantum field theory was originally invented to address the difficult problem of finding diagrammatic descriptions of oscillating integrals with degenerate critical points.  But since then, BV algebras have become interesting objects of study in their own right, and mathematicians sometimes have good understanding of the homological aspects of the story without any access to the diagrammatics.  In this note we reverse the usual direction of argument: we begin by asking for an explicit calculation of the homology of a BV algebra, and from it derive Wick's Theorem and the other Feynman rules for finite-dimensional integrals. 
\end{abstract}

\maketitle


\setcounter{section}{-1}

\begin{pref}
We wrote this note in the spring of 2011 as graduate students, 
when we were seeking to understand how one might invent such an elaborate procedure and  found this elementary motivating example.
In the following years, our understanding of the Batalin--Vilkovisky formalism continued to evolve \cite{OwenThesis,MR3406714,GH}. 
To get other perspectives and to find recent applications of the formalism, we recommend ~\cite{SiLi2017,MnevLectures}.
\end{pref}

\section{Introduction}

Anyone with a glancing interest in physics eventually sees Feynman diagrams and is told that they play a key role in quantum field theory, notably in the process of renormalization of path integrals. If a mathematician digs a bit deeper, she learns that Feynman diagrams fit into a method for constructing asymptotic series associated to oscillating integrals. Recall that an oscillating integral is an expression of the form
\[
\int_\RR f(x) e^{\sqrt{-1}g(x)} \, \d x,
\]
with $f$ and $g$ real functions. The adjective ``oscillating" refers to the variation of the exponential term, especially when $|g'|$ is large. In section \ref{motivation}, we will return to this relationship between Feynman diagrams and integration, but our main goal is to provide an alternative story about how to invent Feynman diagrams, rooted in homological algebra. Again, the basic problem has its source in physics. In the process of trying to better understand quantum field theory and string theory, physicists rediscovered many homological tools,
notably the Chevalley--Eilenberg description of equivariant cohomology, a version of which is called the ``BRST formalism'' in the physics literature.
Building on this, the Russian physicists Batalin and Vilkovisky introduced a formalism to tackle supergravity \cite{MR616572,MR726170,MR736479,MR776145}; 
the BV formalism is now widely considered the most powerful approach to quantizing gauge theories.
We focus this article on the simplest examples of the homological problem that appears in the BV formalism. The problem is as follows. 

Fix a field $\KK$ of characteristic $0$ and a positive integer $N$.  Consider the graded-commutative algebra 
\[
V_\bullet = \KK\llbracket x_1,\dots,x_N,\xi_1,\dots,\xi_N,\hbar\rrbracket
\] 
with its usual topology as a power-series algebra, where $x_1,\dots,x_N,\hbar$ are in degree $0$ (and hence commute with everything), and $\xi_1,\dots,\xi_N$ are in homological degree $1$ (and hence anticommute with each other). Observe that $V_\bullet$ is concentrated between degrees $0$ and $N$, because $\xi_i^2 = 0$.
We make $V_\bullet$ into a chain complex, but {\em not} a differential graded algebra, as follows. Pick a symmetric invertible $N\times N$ matrix $a=(a_{i,j})_{i,j = 1}^N$.  Pick a power series $b(x)\in\KK\llbracket x_1,\dots,x_N\rrbracket \subset V_\bullet$ with only cubic and higher terms.  Equip $V_\bullet$ with the degree-$(-1)$ differential:
\[ 
Q = \sum_{i,j=1}^N a_{i,j}x_i \frac{\partial}{\partial \xi_j} - \sum_{i=1}^N \frac{\partial b(x)}{\partial x_i} \frac{\partial}{\partial \xi_i} - \hbar \sum_{i=1}^N \frac{\partial^2}{\partial x_i \partial \xi_i}. 
\]
These partial derivatives behave as usual so long as one takes into account the appropriate signs. For instance,
\[
\frac{\partial}{\partial \xi_i} \frac{\partial}{\partial \xi_j} = - \frac{\partial}{\partial \xi_j} \frac{\partial}{\partial \xi_i}
\]
because the partial derivative $\frac{\partial}{\partial \xi_i}$ has odd degree. After all, it removes a copy of $\xi_i$ from any monomial $x_1^{m_1} \cdots x_N^{m_N} \xi_1^{n_1} \cdots \xi_N^{n_N}$ and hence lowers the homological degree of the monomial by 1.

\emph{The fundamental goal is to understand the homology of this chain complex and, even more, to know how to explicitly describe the image of closed elements inside the homology.} Let us take as an ansatz that the homology of $(V_\bullet,Q)$ in degree $0$ is isomorphic to $\KK\llbracket \hbar \rrbracket$ as a $\KK\llbracket \hbar \rrbracket$-module.   As $V_\bullet$ is zero in positive degrees, every element in $V_0 = \KK\llbracket x_1,\dots,x_N,\hbar\rrbracket$ is a cycle. The image of $Q$ in $V_0$ is contained in the ideal generated by $x_1,\dots,x_N,\hbar$, since every term in $Q$ either adds a power of $x_i$ or $\hbar$. Thus the number $1 \in V_0 $ is not a boundary.  For $f\in V_0$ we write $[f]$ for its image in homology $H_0(V)$.  According to the ansatz,  for every $f \in V_0$ we should be able to write
\[
[f] = \langle f \rangle [1]
\]
for a unique number $\langle f\rangle \in \KK\llbracket \hbar \rrbracket$.  By succeeding at this, we will verify the ansatz.

The $\langle f \rangle$ are computed by Feynman diagrams.  We will warm up with two examples, and then explain in general how the diagrammatic description arises naturally from the homology calculation.  The chain complex $(V_\bullet,Q)$ is an example of a \define{BV algebra}, and we hope that this example can provide the reader with more intuition for the general BV formalism in perturbative quantum field theory.  We end this note with some general discussion motivating the BV approach in finite-dimensional non-gauged integrals.  We expect that this note contains no results that are not known to the experts, but we hope that it gives an entry point for non-experts trying to learn BV theory.  For other parts of this story, see for example \cite{ABF,Fiorenza04,Stasheff:fk,WittenAntibracket}.

\section{Example: Wick's lemma}

As our first warm-up, we set $N=1$ and $b=0$.  Then our complex is
\[
\begin{array}{ccc}
V_1 & \overset{Q}{\longrightarrow} &  V_0 \\
\shortparallel & & \shortparallel \\
\KK\llbracket x,\hbar \rrbracket \, \xi & & \KK\llbracket x,\hbar \rrbracket
\end{array}
\]
where  
\[ 
Q = ax \frac{\partial}{\partial \xi} - \hbar \frac{\partial^2}{\partial x \partial \xi}. 
\]
Given $\xi\, f(x) \in V_1$, $Q(\xi f) = axf(x) - \hbar f'(x)$. A little formal calculus tells us that $Q(\xi f) = 0$ only when $f(x) = \exp( ax^2/2\hbar)$, but as this $f$ is not in $\KK\llbracket x,\hbar \rrbracket$, the homology in degree $1$ vanishes.

The boundaries comprise the closed $\KK\llbracket h\rrbracket$-span of the image of $Q$ on the monomials $\xi x^n$, where ``closed" means that we take the closure in the power series topology. The base case is $Q(\xi) = ax$. For $n > 0$, we have the elements $Q(\xi x^n) = ax^{n+1} - \hbar n x^{n-1}$. Hence in homology we have $[x^{n+1}] = \frac{\hbar}{a}n [x^{n-1}]$. By recursion $\langle x^{2n+1}\rangle = 0$ and
\[ 
\langle x^{2n}\rangle = \left(\frac\hbar a \right)^n (2n-1)!!,
\]
where 
\[
(2n-1)!!  = (2n-1)(2n-3)\cdots (3)(1) = \frac{(2n)!}{2^n n!}.
\]
This classic result about the moments of a Gaussian distribution is known as \define{Wick's formula} by physicists \cite{Wick}. The combinatorial interpretation is that $(2n-1)!!$ counts the number of pairings of $2n$ objects; each of the $n$ connections in such a pairing should then be understood as contributing a factor of $\hbar/a$.
(Note that we have confirmed the ansatz for $H_0(V_\bullet,Q)$ in this case.)

\begin{xca}
We encourage the reader to pause from this note long enough to consider: what happens when $N > 1$ and $b=0$?
\end{xca}

\section{Example: Counting trivalent graphs}

We consider now the situation that $N=1$, $a=1$, and $b(x) = x^3/6$. The boundaries are spanned by the images of monomials
\[
Q(x^n\xi) =  x^{n+1} - x^{n+2}/2 - \hbar nx^{n-1},
\] 
where we interpret $x^{-1}$ as $0$ to include the case $n = 0$ conveniently. Hence
\[ 
[x^{n+1}] = \frac12 \, [x^{n+2}] + \hbar \, n \, [x^{n-1}]. 
\]
One can thus begin substituting $[x] = \frac12 [x^2] = \frac12\bigl( \frac12 [x^3] + \hbar [1] \bigr) = \frac12\bigl( \frac12 \bigl( \frac12[x^4] + 2\hbar [x] \bigr) + \hbar [1] \bigr) = \dots$. At each stage one adds terms in either high degree in $x$ or high degree in $\hbar$, and so the subsequent infinite sum converges to something in $\KK\llbracket \hbar \rrbracket \cdot [1]$.  But if we use this {\it ad hoc} approach to the substitutions, it becomes combinatorially difficult to compute the coefficients in $\langle x^n \rangle = [x^n]/[1] \in \KK\llbracket \hbar \rrbracket$.

Instead, we posit the following answer, and check that it satisfies the necessary conditions.  A \define{Feynman diagram for $\langle x^n \rangle$} is a connected finite graph with all vertices trivalent, except for one marked vertex that is $n$-valent with totally-ordered incident half-edges (self-loops and parallel edges are allowed).  An \define{automorphism} of a Feynman diagram for $\langle x^n\rangle$ is a permutation of the half-edges of the graph which does not change the data of which half-edges are part of the same edge nor the data of which half edges are incident on a vertex; the permutation also should act trivially on the half edges incident to the marked vertex.
If $\Gamma$ is a Feynman diagram for $\langle x^n \rangle$, its \define{first Betti number} $\beta(\Gamma)$  is the number of (full) edges minus the number of non-marked vertices --- an easy calculation shows that $\beta(\Gamma) = (v(\Gamma) + n)/2$, where $v(\Gamma)$ is the number of unmarked vertices in $\Gamma$.

We list the Feynman diagrams for $\langle x^2 \rangle$ with Betti number $1$ or $2$, along with their numbers of automorphisms:
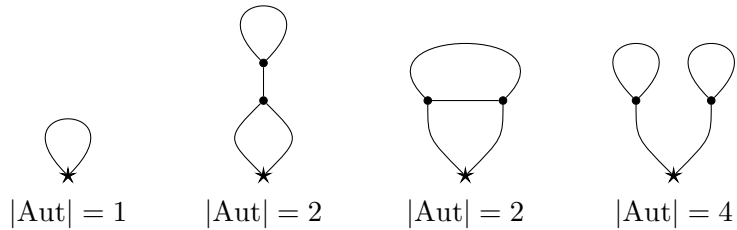
\[
  \begin{tikzpicture}[baseline=(aut.base)]
     \path node[Sdot] (star) {} +(-.5,-.5) coordinate (clip1) +(.5,1) coordinate (clip2) ++(0,-.5) node (aut) {$\lvert \Aut \rvert = 1$} 
      ;
    \clip (clip1) rectangle (clip2);
    \draw (star) .. controls +(-1,1) and +(1,1) .. (star);
  \end{tikzpicture}
  \quad \quad
  \begin{tikzpicture}[baseline=(aut.base)]
    \path node[Sdot] (star) {} +(-.5,-.5) coordinate (clip1) +(.5,2.5) coordinate (clip2) ++(0,-.5) node (aut) {$\lvert \Aut \rvert = 2$}
      ;
    \clip (clip1) rectangle (clip2);
    \path (star) ++(0,1) node[dot] (dot1) {} ++(0,.5) node[dot] (dot2) {};
    \draw (star) .. controls +(-.5,.5) and +(-.5,-.5) .. (dot1);
    \draw (star) .. controls +(.5,.5) and +(.5,-.5) .. (dot1);
    \draw (dot1) -- (dot2);
    \draw (dot2) .. controls +(-1,1) and +(1,1) .. (dot2);
  \end{tikzpicture}
  \quad \quad
  \begin{tikzpicture}[baseline=(aut.base)]
    \path node[Sdot] (star) {} +(-1,-.5) coordinate (clip1) +(1,2.5) coordinate (clip2) ++(0,-.5) node (aut) {$\lvert \Aut \rvert = 2$}
      ;
    \clip (clip1) rectangle (clip2);
    \path (star) +(-.5,1) node[dot] (dot1) {} +(.5,1) node[dot] (dot2) {};
    \draw (star) .. controls +(-.5,.5) and +(0,-.5) .. (dot1);
    \draw (star) .. controls +(.5,.5) and +(0,-.5) .. (dot2);
    \draw (dot1) -- (dot2);
    \draw (dot1) .. controls +(-1,1) and +(1,1) .. (dot2);
  \end{tikzpicture}  
  \quad \quad
  \begin{tikzpicture}[baseline=(aut.base)]
    \path node[Sdot] (star) {} +(-1,-.5) coordinate (clip1) +(1,2.5) coordinate (clip2) ++(0,-.5) node (aut) {$\lvert \Aut \rvert = 4$}
      ;
    \clip (clip1) rectangle (clip2);
    \path (star) +(-.5,1) node[dot] (dot1) {} +(.5,1) node[dot] (dot2) {};
    \draw (star) .. controls +(-.5,.5) and +(0,-.5) .. (dot1);
    \draw (star) .. controls +(.5,.5) and +(0,-.5) .. (dot2);
    \draw (dot1) .. controls +(-1,1) and +(1,1) .. (dot1);
    \draw (dot2) .. controls +(-1,1) and +(1,1) .. (dot2);
  \end{tikzpicture}
\]

We claim that $\langle x^n \rangle = c_n$, where $c_n$ is the following Feynman-style sum:
\[ c_n = \sum_{\substack{\Gamma \text{ a Feynman} \\ \text{diagram for } \langle x^n \rangle}} \frac{\hbar^{\beta(\Gamma)}}{\lvert \Aut \Gamma \rvert}.\]
In particular, $c_0 = 1 = \langle 1 \rangle$ as it should be, and $c_n \in \hbar^{n-1}\KK^n$.  
It therefore suffices to verify that the $c_n$s satisfy the recursion relation
$$ c_{n+1} = \frac12  c_{n+2} + \hbar  n  c_{n-1}, \quad n\geq 0;$$
we set $c_{-1} = 0$ to get things started.
Indeed, the {\it ad hoc} substitutions above imply that this recursion relation has at most one solution with $c_0=1$ such that $c_n \to 0$ in the power series topology on~$\KK\llbracket\hbar\rrbracket$. 

We verify the recursion relation as follows.  Let $\Gamma$ be a Feynman diagram for $\langle x^{n+1} \rangle$.  The half-edges ending on the marked vertex are totally ordered by $1,2,\dots,n+1$. Start walking along the last half-edge.  You will either arrive back at the marked vertex along half-edge number $j$, or you will arrive at a trivalent vertex.  In the first case, delete this loop, thereby creating a Feynman diagram for $\langle x^{n-1} \rangle$. In the second case, you can ``unzip'' $\Gamma$ along half-edge number $n+1$, between the marked vertex and the trivalent vertex. (That is, you slide the point down the half-edge to the starred vertex, much as you slide a zipper down a jacket.) You thereby create a Feynman diagram for $\langle x^{n+2}\rangle$. 
\[
\begin{array}{ccc}
  \begin{tikzpicture}[baseline=(center)]
    \path node[Sdot] (star) {} +(0,.5) coordinate (center) +(.125,.5) node {\scriptsize \dots} +(-1.25,-.1) coordinate (clip1) +(1.25,1.25) coordinate (clip2);
    \clip (clip1) rectangle (clip2);
    \draw (star) -- node[anchor=north east,inner sep=0,near end] {$\scriptstyle 1$} ++(-1,1);
    \draw (star) -- node[anchor=north east,inner sep=0,near end] {$\scriptstyle 2$} ++(-.5,1);
    \draw (star) --  ++(1,1);
    \draw (star) .. controls +(2,1) and +(0,2) .. node[anchor=south east,inner sep=0,very near end] {$\scriptstyle j$} node[anchor=north west,inner sep=0,very near start] {$\scriptstyle n+1$}  (star);
  \end{tikzpicture}  
  \leadsto
  \begin{tikzpicture}[baseline=(center)]
    \path node[Sdot] (star) {} +(0,.5) coordinate (center) +(.125,.5) node {\scriptsize \dots};
    \draw (star) -- node[anchor=north east,inner sep=0,near end] {$\scriptstyle 1$} ++(-1,1);
    \draw (star) -- node[anchor=north east,inner sep=0,near end] {$\scriptstyle 2$} ++(-.5,1);
    \draw (star) --  ++(1,1);
  \end{tikzpicture}  
  & \quad\quad\quad\quad &  
    \begin{tikzpicture}[baseline=(center)]
    \path node[Sdot] (star) {} +(0,.5) coordinate (center) +(.125,.5) node {\scriptsize \dots};
    \draw (star) -- node[anchor=north east,inner sep=0,near end] {$\scriptstyle 1$} ++(-1,1);
    \draw (star) -- node[anchor=north east,inner sep=0,near end] {$\scriptstyle 2$} ++(-.5,1);
    \draw (star) -- node[anchor=north west,inner sep=0] {$\scriptstyle n+1$} ++(.75,.75) node[dot] (dot) {};
    \draw (dot) -- ++(-.25,.5);
    \draw (dot) -- ++(.25,.5);
  \end{tikzpicture}
  \leadsto
  \begin{tikzpicture}[baseline=(center)]
    \path node[Sdot] (star) {} +(0,.5) coordinate (center) +(0,.75) node {\scriptsize \dots};
    \draw (star) -- node[anchor=north east,inner sep=0,near end] {$\scriptstyle 1$} ++(-1,1);
    \draw (star) -- node[anchor=north east,inner sep=0,near end] {$\scriptstyle 2$} ++(-.5,1);
    \draw (star) -- node[anchor=south east,inner sep=0,near end] {$\scriptstyle n+1$} ++(.5,1.25);
    \draw (star) -- node[anchor=north west,inner sep=0,near end] {$\scriptstyle n+2$} ++(1,1.25);
  \end{tikzpicture} \\
    \text{Delete} & & \text{Unzip}
  \end{array}
\]
Every diagram for $\langle x^{n+2}\rangle$ or $\langle x^{n-1} \rangle$ can be created this way from some diagram for $\langle x^{n+1} \rangle$. To reverse the process, either zip up the last two edges in a diagram for $\langle x^{n+2}\rangle$, or add a self-loop to the marked vertex in a diagram for $\langle x^{n-1}\rangle$.  

We must now count, with symmetry, how many times each diagram is created.  First, we consider the unzipping operation.  If our initial diagram $\Gamma$ for $\langle x^{n+1} \rangle$ had a symmetry switching the two half-edges on that first trivalent vertex, then unzipping broke this symmetry, and so divided the automorphism group by $2$.  If it did not have such a symmetry, then in fact there were two distinct ways to unzip $\Gamma$ to get the same diagram for $\langle x^{n+2}\rangle$, because we did not say which edge should become number $n+1$ and which should become number $n+2$.  Together, these are the factor of $\frac12$ in $c_{n+1} = \frac12 c_{n+2} + \dots$.  Second, we have the case of deleting a self-loop: for a given diagram for $\langle x^{n-1}\rangle$, there are precisely $n$ ways to add a self-loop to the marked vertex such that one end of the self-loop is half-edge number $n+1$, hence the factor of $n$ in $c_{n+1} = \dots + \hbar n c_{n-1}$.  The unzipping operation does not change the Betti number of a diagram, but deleting a self-loop does, hence the factors of $\hbar$.  This completes the proof.

\begin{xca}
  What happens when $N=1$, $a=1$, and $b(x) = \frac{x^4}{4!}$?  What about when $b(x) = \frac{x^3}6 + \frac{x^4}{4!}$?
\end{xca}

\begin{remark}  
There is a different solution to the recursion relation,
which highlights some interesting issues about convergence.
Define the following sequence in $\KK\llbracket \hbar \rrbracket$:
\[ d_n = 12^n \sum_{k = \lceil \frac n 2 \rceil}^\infty \left( \frac\hbar{288}\right)^k \frac{(6k-2n)!}{(3k-n)!\,(2k-n)!} \]
These $d_n$ also satisfy the recursion relation, and $d_0 = 1 + \frac 5 {24} \hbar + \dots$ is invertible, so $c_n = d_n/d_0$.  When $\KK=\RR$, one can estimate the growth rates of the coefficients of $d_n$: $\frac{(6k-2n)!}{(3k-n)!\,(2k-n)!} \sim 6^k k!$, and so $d_n \in \RR \llbracket \hbar \rrbracket$ has zero radius of convergence.  In fact, the power series $c_n$ also have zero radius of convergence for $n\geq 1$, because their coefficients in $\hbar$ grow roughly as the coefficients of $\log d_0$.  For example, one can show that $d_1 = 3\frac{\partial d_0}{\partial\hbar}$, and so $c_1 = 3\frac{\partial}{\partial\hbar}\log d_0$, so if $c_1$ were the Taylor expansion at $0$ of something analytic in $\hbar$, then $d_0$ would also have positive radius of convergence.
(On the other hand, the series $c_n$ do have positive radius of convergence over $\QQ_p$ for $p>0$.)
\end{remark}

\section{The general case: From homological algebra to diagrammatics}

We now consider the general case.  Recall that we fix $N$ a positive integer and $\KK$ a field of characteristic $0$, and we build the graded commutative algebra $V_\bullet = \KK\llbracket x_1,\dots,x_N,\xi_1,\dots,\xi_N,\hbar \rrbracket$ as a completed symmetric algebra, with the generators are $x_1,\dots,x_N,\hbar$ in grading zero and $\xi_1,\dots,\xi_N$ in grading $1$.  We then choose an invertible symmetric $N\times N$ matrix $(a_{ij})$ and a power series $b(x) \in \KK\llbracket x_1,\dots,x_N \rrbracket$ that vanishes at least to order $3$.  With this data, we make $V_\bullet$ into a chain complex by choosing the differential:
\[ Q = \sum_{i,j=1}^N a_{ij}x_i \frac{\partial}{\partial \xi_j} - \sum_{i=1}^N \frac{\partial b(x)}{\partial x_i} \frac{\partial}{\partial \xi_i} - \hbar \sum_{i=1}^N \frac{\partial^2}{\partial x_i \partial \xi_i}. \]
For $f \in \KK\llbracket x_1,\dots,x_N,\hbar \rrbracket$, we write $[f]$ for its image in the $Q$-homology of $V_\bullet$.  In light of the ansatz that $H_0(V_\bullet,Q) = \KK\llbracket \hbar \rrbracket$, we are interested in computing $\langle f \rangle = [f]/[1] \in \KK\llbracket \hbar \rrbracket$.  It suffices to compute $\langle \sum_{i_1,\dots,i_n = 1}^N f_{i_1,\ldots,i_n}x_{i_1}\cdots x_{i_n} \rangle$ for each $n$-tensor $(f_{i_1,\ldots,i_n})_{\vec\imath \in \{1,\dots,N\}^n} \in \KK^{N^n}$.

One approach is to tackle the algebra directly. We (temporarily) adopt the following index-full notation.  For $\vec \imath \in \{1,\dots,N\}^m$, we write $x_{\vec\imath}$ for $x_{i_1}\cdots x_{i_m} \in \KK[x_1,\dots,x_N]$, so that for example the $m$fold power of $x_1$ is written $(x_1)^m = x_{{\scriptstyle 1,\dots,1}} 
$.  We define the Taylor coefficients of $b$ via $b^{(m)}_{\vec \imath} = \frac{\partial^m b}{\partial x_{i_1}\cdots \partial x_{i_m}} \bigr|_{0}$.  In particular, each $b^{(m)}$ is a symmetric $m$-tensor, and
$ \frac{\partial b(x)}{\partial x_i} = \sum_{m=2}^\infty \frac1{m!} \sum_{\vec \jmath \in \{1,\dots,N\}^m} b^{(m+1)}_{i,\vec \jmath} x_{\vec \jmath}\,. $

For each $(n+1)$-tensor $(f_{i,\vec \imath})_{{i\in \{1,\dots,N\},\,\vec\imath \in \{1,\dots,N\}^n}}$, consider the element $$\sum_{i,\vec \imath,j}  f_{i,\vec \imath} \,x_{\vec \imath} \,(a^{-1})_{i,j} \,\xi_j \in V_1,$$ where $a^{-1}$ is the matrix inverse to $(a_{i,j})$.  These elements, running over all such tensors, span $V_1$ (rather, for the power-series topology, their span is dense in $V_1$).  Thus, the image under $Q$ of this basis spans the boundaries in $V_0$ (in the power-series topology):
\begin{multline*}
  Q\left( \sum_{i,\vec\imath} f_{i,\vec \imath} \,x_{\vec \imath} \,(a^{-1})_{i,j} \,\xi_j\right)  \\
  = \sum_{i,\vec\imath} f_{i,\vec \imath} \, x_i x_{\vec \imath}
   - \sum_{m=2}^\infty \sum_{i,\vec\imath,j,\vec\jmath} \frac1{m!} \,b^{(m+1)}_{j,\vec\jmath}\, x_{\vec\jmath} \, \, (a^{-1})_{i,j} \, f_{i,\vec\imath}\, x_{\vec\imath}
   \\  - \hbar \sum_{i,\vec\jmath} \sum_{k=1}^n f_{i,\vec \jmath}\, (a^{-1})_{i,j_k} x_{j_1,\dots,\hat \jmath_k,\dots,j_n}.
\end{multline*}
By ``$\hat \jmath_k$'' we mean ``remove this term from the list.''  Thus we can write the class $\left[\sum_{i,\vec\imath} f_{i,\vec \imath} \, x_i x_{\vec \imath}\right]$ as a sum of various other terms, each of which has either more $x$s or more $\hbar$s.

This index-full notation is, of course, a mess.  Much better is a Feynman-diagrammatic notation generalized by Penrose to handle contractions of tensors \cite{Penrose1971}.  We define a \define{Feynman diagram} to be a finite connected graph (self-loops and parallel edges are allowed) built from the following pieces:
\begin{itemize}
  \item Precisely one \define{marked} vertex, with valence $n$, which is labeled by an $n$-tensor $f \in \KK^{N^n}$, and whose incident half-edges are totally ordered; we will draw the marked vertex with a star \tikz[baseline=(bottom)] \node[Sdot] (star) {} ++(0,-.7ex) coordinate (bottom);, and leave the tensor and the total ordering implicit.
  \item Some number of \define{internal} vertices, which are required to have valence $3$ or more; we will draw internal vertices as solid bullets \tikz[baseline=(bottom)] \node[dot] (star) {} ++(0,-.7ex) coordinate (bottom);.
  \item Some number of univalent \define{external} vertices; we will draw external vertices as open circles~\tikz[baseline=(bottom)] \node[Odot] (star) {} ++(0,-.7ex) coordinate (bottom);.
\end{itemize}
An \define{automorphism} of a Feynman diagram is a permutation of its half-edges that does not change the combinatorial type of the diagram --- it may separately permute both the internal and external vertices, but it should not permute the half-edges incident to the marked vertex.  Given a Feynman diagram $\Gamma$, its \define{first Betti number} $\beta(\Gamma)$ is its total number of edges minus its number of un-marked vertices.  We say that an edge is \define{internal} if it connects internal and marked vertices, and \define{external} if one of its ends is an external vertex.

We now define the \define{evaluation} $\ev(\Gamma)$ of a Feynman diagram $\Gamma$ as follows.  First, suppose we are given a labeling of the half-edges by numbers $\{1,\dots,N\}$.  To such a labeled Feynman diagram we associate a product of matrix coefficients:
\begin{itemize}
  \item The marked vertex contributes $f_{\vec\imath}$, where $\vec \imath$ is the vector of labels formed by reading the labels on the incident half-edges in the prescribed order (recall that part of the data of $\Gamma$ was a total ordering of these vertices).
  \item Each internal vertex with valence $m$ contributes $b^{(m)}_{\vec \imath}$, where $\vec \imath$ is the vector of labels formed by reading the incident half-edges in any order (recall that the tensors $b^{(m)}$ are symmetric).
  \item Each external vertex with incident half-edge labeled by $i \in \{1,\dots,N\}$ contributes the variable $x_i \in V_0 = \KK\llbracket x_1,\dots,x_N,\hbar \rrbracket$.
  \item Each internal edge with half-edges labeled $i,j$ contributes $(a^{-1})_{i,j} = (a^{-1})_{j,i}$.
  \item Each external edge with half-edges labeled $i,j$ contributes $\delta_{i,j}$.
\end{itemize}
Thus a labeled Feynman diagram evaluates to some monomial in $V_0$.  The evaluation $\ev(\Gamma)$ of an unlabeled Feynman diagram $\Gamma$ is defined to be the sum over all possible labelings of its evaluation as a labeled Feynman diagram.  Finally, we give a map $\{\text{Feynman diagrams} \} \to V_0$ by: 
\[ \Gamma \mapsto \frac{\ev(\Gamma)\hbar^{\beta(\Gamma)}}{\lvert \Aut(\Gamma)\rvert}. \]

In this notation, our index-full calculation above was the statement that, for a fixed tensor $f$ labeling the marked vertices:\vspace{-1ex}

\noindent\parbox{\textwidth}{\[
  \begin{tikzpicture}[baseline=(center)]
    \path node[Sdot] (star) {} +(0,.5) coordinate (center) +(0,.75) node {\scriptsize \dots};
    \path (star) +(-.75,1) node[Odot] (dot1) {}
      +(-.5,1) node[Odot] (dot2) {}
      +(.5,1) node[Odot] (dot3) {}
      +(.75,1) node[Odot] (dot4) {};
    \draw (star) -- (dot1); \draw (star) -- (dot2); \draw (star) -- (dot3); \draw (star) -- (dot4);
    \path (dot1) ++(-.125,.125) coordinate (b1) (dot4) ++(.125,.125) coordinate (b2);
    \draw[decorate,decoration=brace] (b1) -- node[auto] {$\scriptstyle n+1$} (b2);
  \end{tikzpicture}
  -
  \sum_{m=2}^\infty
  \begin{tikzpicture}[baseline=(center)]
    \path node[Sdot] (star) {} +(0,.5) coordinate (center);
    \path (star) +(-.75,1) node[Odot] (dot1) {}
      +(-.5,1) node[Odot] (dot2) {}
      +(.5,1) node[Odot] (dot3) {}
      +(0,.75) node {\scriptsize \dots}
      ++(1.25,0) node[dot] (dot) {}
        +(-.375,1) node[Odot] (dot4) {}
        +(.375,1) node[Odot] (dot5) {}
        +(0,.75) node {\scriptsize \dots};
    \draw (star) -- (dot1); \draw (star) -- (dot2); \draw (star) -- (dot3); 
    \draw (star) ..controls +(.5,.5) and +(-.5,.5).. (dot);
    \draw (dot) -- (dot4); \draw (dot) -- (dot5);
    \path (dot1) ++(-.125,.125) coordinate (b1) (dot3) ++(.125,.125) coordinate (b2);
    \draw[decorate,decoration=brace] (b1) -- node[auto] {$\scriptstyle n$} (b2);
    \path (dot4) ++(-.125,.125) coordinate (b3) (dot5) ++(.125,.125) coordinate (b4);
    \draw[decorate,decoration=brace] (b3) -- node[auto] {$\scriptstyle m$} (b4);
  \end{tikzpicture}
  -
  \sum_{k=1}^n
  \begin{tikzpicture}[baseline=(center)]
    \path node[Sdot] (star) {} +(0,.5) coordinate (center) +(0,1) node {\scriptsize \dots};
    \path (star) +(-.75,1) node[Odot] (dot1) {}
      +(-.5,1) node[Odot] (dot2) {}
      +(.5,1) node[Odot] (dot3) {};
    \draw (star) -- (dot1); \draw (star) -- (dot2); \draw (star) -- (dot3);
    \path (dot1) ++(-.125,.125) coordinate (b1) (dot3) ++(.125,.125) coordinate (b2);
    \draw[decorate,decoration=brace] (b1) -- node[auto] {$\scriptstyle n-1$} (b2);
    \draw (star) .. controls +(.75,.75) and +(-.25,1) .. node[anchor=south east,inner sep=0, near end] {$\scriptstyle k$}  (star);
  \end{tikzpicture}
  \quad\text{ is a boundary in }V_0.
\]
In the final diagram, the self-loop connects the $k$th and $(n+1)$th half-edges on the marked vertex.}

We can therefore evaluate any $\bigl\langle \sum_{\vec\imath}\, f_{\vec\imath}\, x_{\vec\imath} \bigr\rangle$, and indeed $\bigl\langle \frac{\ev(\Gamma)\hbar^{\beta(\Gamma)}}{\lvert \Aut(\Gamma)\rvert}\bigr\rangle$ for any Feynman diagram $\Gamma$, by playing Hercules' game of the many-headed Hydra, as follows.  Pick some external vertex.  Either we try to chop it off, forming a new internal vertex, at which point our Hydra grows at least two new external vertices (``heads''), or we attach it to some other external vertex, thus forming a loop and  increasing the Betti number of the graph. (This description holds modulo boundaries in $V_\bullet$.) In the power-series topology, any sequence of Hydra with strictly-increasing head number converges to $0$, and for any given $\beta$ the game only produces finitely many graphs with Betti number $\leq \beta$.  Thus the whole game converges in the power-series topology.

What does it converge to?  The only Feynman diagrams left at the end of the game are those with no external vertices at all, since these are the only Hydra that do not have a head that Hercules can chop off.  All together, we have proved that for any tensor $f_{\vec\imath}\,$,
\[ 
\left\langle \sum_{\vec\imath}\, f_{\vec\imath}\, x_{\vec\imath} \right\rangle = \sum_{\substack{\text{Feynman diagrams }\Gamma \\ \text{with no external vertices} \\ \text{and marked vertex labeled by }f}} \frac{\ev(\Gamma)\,\hbar^{\beta(\Gamma)}}{\lvert\Aut(\Gamma)\rvert}  
\]
as a formal power series in $\KK\llbracket \hbar \rrbracket$.

\begin{remark}
These techniques and this result are examples of {\em homological perturbation theory}, which allows one to understand a chain complex as a ``perturbation" of a simpler chain complex.  On the same graded vector space $V_{\bullet} = \KK\llbracket x_{1},\dots,x_{N},\xi_{1},\dots,\xi_{N},\hbar\rrbracket$ we could consider the very simple differential $Q_{0} = \sum_{i,j=1}^N a_{ij}x_i \frac{\partial}{\partial \xi_j}$.  An easy exercise shows that the homology of $V_{\bullet}$ consists of a copy of $\KK\llbracket\hbar\rrbracket$ in degree $0$, and that for $Q_{0}$ the map $\langle\cdot\rangle : V_{\bullet} \to \KK\llbracket\hbar\rrbracket$ is the algebra homomorphism that sets all the $x_{i}$ and $\xi_{j}$ to~$0$.

The differential $Q$ that we actually care about is a ``perturbation'' of $Q_{0}$ in the sense that $Q - Q_{0}$ is ``small'', and, in fact, ``much smaller than $Q_{0}$'' in the power-series topology.  There is a general theory that allows one to analyze (and, indeed, write formulas for) chain complexes which are perturbations of understood complexes.  In particular, whenever one has a perturbation of a chain complex, one can construct a differential $\delta$ on $\H_{\bullet}(V_{\bullet},Q_{0})$, such that the homology of $\bigl( \H_{\bullet}(V_{\bullet},Q_{0}), \delta \bigr)$ is identified with the homology of $(V_{\bullet},Q)$.  But in our case $\H_{\bullet}(V_{\bullet},Q_{0}) = \KK\llbracket \hbar\rrbracket$, and so the differential $\delta$ is necessarily zero.  Thus $\H_{\bullet}(V_{\bullet},Q) = \KK\llbracket \hbar\rrbracket$ as well, justifying our original ansatz from the introduction.  Perturbing the differential does, however, change the map $\langle\cdot\rangle$, and although the perturbed map can be fully described with general homological perturbation theory, in our situation of interest the direct analysis is shorter.  For more details on homological perturbation theory, see~\cite{Crainic04}.
\end{remark}

\section{Motivation: Finite dimensional integrals}\label{motivation}

A different explanation is available for why $\langle f \rangle$ is computable as a sum of Feynman diagrams.  To explain it, we move to a somewhat more general problem, explained in detail in \cite{WittenAntibracket}. In brief, we relate the usual way of encoding integration using homological algebra --- namely, the de Rham complex --- to the homological constructions we've discussed so far.

Let $X$ be an $N$-dimensional smooth manifold over $\RR$.  In analogy with the de Rham complex $\Omega^\bullet = \Gamma( \bigwedge^{\bullet}\T^* X )$, the manifold $X$ determines a graded commutative algebra $V_\bullet = \Gamma( \bigwedge^{\bullet}\T X )$ of antisymmetric multivector fields.  Unlike the de Rham complex, $V_\bullet$ is not canonically a chain complex.  But we can make it into one: when $X$ is oriented, any choice of nowhere-vanishing smooth measure $\mu$ on $X$ determines an isomorphism of graded vector spaces $i_\mu: V_\bullet \isom \Omega^{N - \bullet}$, sending a multivector field $Y_1 \wedge \cdots \wedge Y_k$ to its contraction with $\mu$, which will be an $(N-k)$-form. For instance, when $k = 0$, we send a function $f \in V_0$ to the top form $f \mu \in \Omega^N$. Under this isomorphism of vector spaces, we can transfer the exterior derivative $\d : \Omega^\bullet \to \Omega^{\bullet + 1}$ to define the \define{divergence} operator $\Delta_\mu : V_\bullet \to V_{\bullet - 1}$.  We remark that $\Delta_\mu$ only depends on $\mu$ up to global rescalings: $\Delta_\mu = \Delta_{r \mu}$ for $r\in \RR^\times$.

We call the resulting differential ``$\Delta$'' because it is a second-order differential operator for the algebra structure on $V_\bullet$.  Pick local coordinates $x_1,\dots,x_N$ on $X$ such that $\mu = \lvert \d x_1 \cdots \d x_N\rvert$.  Then locally $V_\bullet$ has generators $x_1,\dots,x_N$ in degree $0$ and $\xi_1,\dots,\xi_N$ in degree $1$, where the generator $\xi_i \in V_1 = \Gamma(\T X)$ is nothing but the vector field $\frac{\partial}{\partial x_i}$. Then $\Delta_\mu = \sum_i\frac{\partial^2}{\partial x_i\partial \xi_i}$. Observe that $(V_\bullet,\Delta_\mu)$ is {\em not} a dg algebra.  Rather, it is an example of a BV algebra. {The notion of a BV algebra axiomatizes the structure implicit in this example of multivector fields, and it allows one to apply the diagrammatics we've developed to other circumstances. The origins of BV algebras lie in quantum field theory. In that case, the ``fields" form an infinite-dimensional manifold. The ordinary de Rham complex is not the right way to obtain volume forms or measures since there are no ``top forms" for an infinite-dimensional manifold. By contrast, the multivector fields still work and thus provide a homological approach to defining volume forms.}

A \define{BV algebra} is a graded-commutative algebra $V_\bullet$ with a degree-$(-1)$ square-zero second-order differential operator $\Delta$. The failure of $\Delta$ to be a derivation equips $V_\bullet$ with a Poisson bracket $\{ -, - \}$ of degree-$(-1)$: we set
\[
\{ f, g \} := \Delta (fg) - \left( (\Delta f) g + (-1)^{|f|} f (\Delta g) \right),
\]
and see that this is a derivation in each slot because $\Delta$ is a second-order differential operator. In our case of the multivector fields, the failure of $\Delta_\mu$ to be a derivation is measured by the {\em Schouten-Nijenhuis bracket}, the usual Lie bracket of vector fields extended to the exterior algebra in the natural way.

Provided $X$ is compact and connected, the $N$th cohomology of $\Omega^\bullet$, and hence the $0$th homology of $V_\bullet$, is one-dimensional.  The map 
\[
[\cdot] : V_0 = \Ccc^\infty(X) \to H_0(V_\bullet) \cong \RR
\] 
is, up to rescaling, precisely the usual integration map $f \mapsto \int_X f \mu$.  In particular, the boundaries in $V_0$ are the functions with total integral $0$, and, defining $\langle f \rangle = [f]/[1]$ as before, we see that $\langle f \rangle$ is precisely the \define{expectation value} $\langle f \rangle = \frac{\int f \mu }{ \int \mu}$.

Pick a function $s \in \Ccc^\infty(X)$, and let $\hbar$ range over $\RR_+$.  Fix a volume form $\mu$ on $X$, and consider the family of measures $\exp(-s/\hbar)\mu$. This produces a family of divergence operators and so of chain complexes, but all the complexes are isomorphic to the de Rham complex on $X$, by construction. It's not difficult to compute the corresponding BV structures:
\[ \Delta_{\exp(-s/\hbar)\mu} = -\frac1\hbar \sum_{i=1}^N \frac{\partial s}{\partial x_i} \frac{\partial}{\partial \xi_i} + \Delta_\mu, \]
where $x_1,\dots,x_N$ are any system of local coordinates on $X$, and $\xi_1,\dots,\xi_N$ are the corresponding degree-$1$ generators of $V_\bullet$.  (In coordinate-free language, the operator $\sum_{i=1}^N \frac{\partial s}{\partial x_i} \frac{\partial}{\partial \xi_i}$ on $V_{\bullet} = \Gamma(\bigwedge^{\bullet}\T X)$ is the operator ``contract with $\d s$.'')  As all the complexes are isomorphic, the homology of $\Delta_{\exp(-s/\hbar)\mu}$ is the same as the homology of $-\hbar \Delta_{\exp(-s/\hbar)\mu} = \sum \frac{\partial s}{\partial x_i} \frac{\partial}{\partial \xi_i} -\hbar\Delta_\mu$.

Let us now consider the case when $s$ has a unique minimum (with invertible Hessian), in the limit as $\hbar \to 0$.  Then the measure $\exp(-s/\hbar)\mu$ is asymptotically supported in an infinitesimal neighborhood of this critical point.  Pick local coordinates $x_1,\dots,x_N$ with the critical point at the origin and satisfying $\mu = \lvert \d x_1 \cdots \d x_N\rvert$.  Then:
\[ -\hbar \Delta_{\exp(-s/\hbar)\mu} = \sum_{i=1}^N \frac{\partial s}{\partial x_i} \frac{\partial}{\partial \xi_i} -\hbar\sum_{i=1}^N \frac{\partial^2}{\partial x_i\partial \xi_i}. \]
The choice of coordinates determines a splitting $s(x) = s(0) + \frac12 \sum_{i,j} a_{i,j} x_ix_j - b(x)$ for $(a_{i,j})_{i,j=1}^N$ a symmetric positive-definite matrix, and $b(x) \in \Ccc^\infty(X)$ vanishing at least cubicly.  Then in terms of Taylor series, $-\hbar \Delta_{\exp(-s/\hbar)\mu} $ is precisely the differential $Q$ that we considered above.  The case when the matrix $(a_{i,j})$ is symmetric and invertible but not positive-definite can be treated similarly, by considering oscillating measures of the form $\exp(\sqrt{-1}s/\hbar)\mu$ rather than the exponentially suppressed measures $\exp(-s/\hbar)\mu$.

Then to calculate the $\hbar\to 0$ asymptotics of expectation values $\langle f \rangle$, we can on the one hand proceed as above with homological calculations, and on the other hand we can study asymptotics of finite-dimensional integrals.  But these asymptotics are well-known to be computed by Feynman diagrams: one simply uses repeated integration by parts to recover the kind of diagrams we have drawn \cite{Dyson1949a,Feynman1950,MR2131010}.  This is the secret reason that the homological problem we began with had a diagrammatic answer.

\begin{remark}
When $\pi_1(X) \neq 1$, not all BV algebra structures compatible with the Schouten-Nijenhuis bracket arise from measures.  When $X$ is connected, the data of a nowhere-vanishing measure up to global rescaling is the same as a flat, holonomy-free connection on the density line bundle over $X$.  In general, BV algebra structures on $V_\bullet$ are in bijection with flat connections on the density line bundle, possibly with nontrivial holonomy \cite{Koszul-classifyBV}.  Among other lessons of the BV formalism is that it makes sense to talk about ``expectation values'' for this more general kind of ``measure.''
\end{remark}

\bibliography{BV}{}
\bibliographystyle{amsalpha}

\end{document}